# PlanTUS:

# A heuristic tool for prospective planning of transcranial ultrasound transducer placements

Maximilian Lueckel[1,2], Suhas Vijayakumar[2], Til Ole Bergmann[1,2]

[1] Leibniz Institute for Resilience Research (LIR) gGmbH, Mainz, Germany

[2] Neuroimaging Center (NIC), Focus Program Translational Neuroscience (FTN), Johannes Gutenberg University Medical Center, Mainz, Germany

**Abstract**

Low-intensity focused transcranial ultrasonic stimulation (TUS) offers unique depth and precision in non-invasive brain stimulation. Effective administration of TUS, however, requires precise placement of transducers to ensure selective neural target exposure and engagement. This process is constrained by individual skull anatomy, hardware limitations, as well as computational and time costs of running exhaustive acoustic simulations to optimize transducer placement. To address these challenges, we introduce PlanTUS, a fast, open-source tool designed to heuristically identify feasible transducer positions, accounting for individual anatomical and hardware constraints. It visualizes relevant metrics on the scalp, such as target accessibility, required transducer tilt, and skull thickness, allowing users to identify and export potential transducer positions compatible with both acoustic simulation and neuronavigation software. In addition to individualized planning, PlanTUS facilitates feasibility evaluations during study planning, offering practical utility to researchers seeking to optimize TUS delivery with greater efficiency and precision.

# 1 Introduction

Low-intensity focused transcranial ultrasonic stimulation (TUS) is a novel non-invasive brain stimulation technique that promises unprecedented depth and focality of stimulation. It utilizes the mechanical energy of ultrasound waves, generated by specialized transducers that are placed on a person's scalp (Darmani et al., 2022). Where and how exactly a transducer is positioned strongly determines ultrasound transmission into the brain and, thus, target exposure and neuronal target engagement. To ensure sufficient and selective overlap of the acoustic focus with the target structure (i.e., to maximize on- and minimize off-target stimulation), individual optimization of transducer placement(s) is essential (Murphy et al., 2025).

The selection of transducer placements is, however, far from trivial and depends on multiple factors, including: (i) the individual anatomy of the skull (which reflects, attenuates, and aberrates the ultrasound, depending on its shape, thickness, and composition), and brain (e.g., shape, size, and position of the target structure); (ii) the technical constraints of TUS systems currently used in human research and clinical applications (particularly single-element or annular array systems with limited focal depths and/or without (lateral) focus steering or bone aberration correction); (iii) available options for transducer coupling (e.g., gel pads, water balloons, etc.) (Murphy et al., 2025).

Accounting for (many of) these factors, acoustic simulations are the current gold standard for evaluating the suitability of transducer placements, by estimating the likely location of the resulting ultrasound focus within the brain (Angla et al., 2023). At the same time, acoustic simulations are often computationally demanding and time-consuming, rendering, e.g., exhaustive search strategies for placement optimization infeasible. Even though accelerated (real-time) simulation solutions (e.g., Choi et al., 2022; Park et al., 2023; Naftchi-Ardebili et al., 2024) and optimization methods for transducer placement (e.g., Park et al., 2019, 2022; Atkinson-Clement & Kaiser, 2024) have been proposed, they often remain computationally expensive (e.g., require model training), consider only a subset of relevant factors (e.g., transducer angulation), or are too complex for non-experts.

We therefore developed PlanTUS – an easy-to-use, open-source tool that quickly provides practical and intuitive, heuristic information about where the TUS transducer can, should, and must not be placed on the scalp of a given individual for targeting a specific brain region of interest. PlanTUS thereby helps to narrow down the large number of potential transducer placements to the most promising one(s), which then can (and need to) be validated with proper acoustic simulations in a feasible amount of time. Importantly, PlanTUS outputs are compatible with different acoustic simulation and neuronavigation software, making it a versatile tool for (individual) prospective planning of transducer placements.

## 2 Methods

### 2.1 Inputs

*Person-specific inputs.* PlanTUS requires a T1-weighted MR image of the individual's head and a mask of the target region in the same space. Using the Charm pipeline (Puonti et al., 2020; provided as part of SimNIBS; Thielscher et al., 2015), the anatomical image is segmented into different tissue types, from which PlanTUS extracts 3D models of the skin and skull.

*Transducer-specific inputs.* The user specifies (i) the minimum and maximum focal depth of the transducer, (ii) its aperture diameter, (iii) the maximum accepted transducer tilt relative to the head surface, and (iv) potential offsets between the transducer's radiating surface and the exit plane as well as (v) between the person's scalp and the transducer surface (e.g., due to coupling media or hair). In addition, (vi) a list of possible focal distances and the corresponding FLHM values of the resulting focus (e.g., according to calibration reports by the manufacturer or own hydrophone acoustic measurements) is required.

### 2.2 Outputs

Based on these inputs, PlanTUS computes and visualizes (using the Connectome Workbench viewer; Marcus et al., 2013) several useful metrics for each potential transducer position (i.e., each point) on the scalp surface that help users to intuitively evaluate potential transducer positions (Figure 1A).

These metrics comprise (i) the distance between each point on the skin and the center of the target brain region (with a black outline indicating the restricted area on the head surface from which the target is reachable, given the limited focal depth of the transducer); (ii) the intersection between the target brain region and an idealized straight trajectory of the ultrasound beam pointing perpendicularly from the skin surface into the brain; (iii) the angle of transducer tilt relative to the skin surface required to make the idealized beam trajectory intersect with the target center; and (iv) the angular deviation between skin and underlying skull surface. Note that PlanTUS also automatically identifies avoidance regions (grey areas on the skin surface in Figures 1 and 2) where placing a transducer is either practically unfeasible (e.g., around the ears, eyes, or nose) or useless (e.g., above and around air-filled cavities/sinuses). In addition, skull thickness is computed and visualized (Figure 1A).

PlanTUS outputs several files that can be used as inputs for external acoustic simulation software (e.g., BabelBrain (Pichardo, 2023), k-wave (Bradley E. Treeby et al., 2010), or k-Plan (https://k-plan.io/)) as well as neuronavigation software (e.g., Localite (https://www.localite.de/en/home/) or BrainSight (https://www.rogue-research.com/tms/brainsight-tms/)). These outputs are crucial for simulation-based validation of the heuristically selected transducer placement(s) and MR-guided navigation of the transducer to the selected position(s), respectively.

*2.3 Code availability*

PlanTUS is fully based on open-source software (Python, SimNIBS (Thielscher et al., 2015), FSL (Jenkinson et al., 2012), FreeSurfer (Fischl, 2012), Connectome Workbench (Marcus et al., 2013)). Code and further instructions are available at: https://plan-tus.org.

# 3 Results

## 3.1 Example use case

Figure 1 illustrates the individual planning of a transducer placement for sonicating the left nucleus accumbens (NAcc) using a 4-channel annular array transducer with a maximum focal depth of 76.6 mm (NeuroFUS CTX-500-4, Brainbox Ltd., Cardiff, UK). Figure 1A shows the main interface that appears after running PlanTUS, with the heuristic metrics visualized on the individual's head surface. To select a transducer placement, users can simply click wherever they want to place the center of the transducer, marked by a small white sphere. Note that in the given example, the transducer was placed at an ideal spot that (i) is close to the target (i.e., within the reach of the transducer-specific focal distance), (ii) has a strong intersection of the idealized acoustic beam trajectory with the target and therefore (iii) does not require a (large) transducer tilt, and (iv) has a similar curvature as the underlying skull bone. The oblique volume (i.e., relative to the transducer axes) view allows to preliminarily check the intersection between the target region and the idealized acoustic beam trajectory (blue/green straight line) pointing from a given position into the brain.

After selecting a position, a new window shows the resulting transducer placement and a simplified representation of the expected acoustic focus overlaid on the target mask and anatomical MR image (Figure 1B, left). The oblique volume view helps users to evaluate the expected on- vs. off-target stimulation in terms of overlap between the simplified acoustic focus and the target volume (Figure 1B, right).

Figure 1C shows an example for the use of PlanTUS outputs (i.e., selected transducer placement(s)) in external software for acoustic simulations (Figure 1C, left; here: k-Plan) and neuronavigation (Figure 1C, right; here: Localite).

Eventually, acoustic simulation results (e.g., acoustic pressure maps) can be evaluated in the same environment (Figure 1D, left) after loading the respective result files. Again, overlap between the simulated acoustic focus and target region (i.e., on- vs. off-target stimulation) can be evaluated using the oblique volume view (Figure 1D, right).

In some cases it may be helpful to select and evaluate a number of different transducer positions with comparable characteristics (e.g., when the ideal position is potentially unfeasible because of experimental setup constraints such as headphones or transducer fixation equipment) or even with different trade-offs between criteria (e.g., overlap between focus and target vs. angulation relative to skull bone and thus attenuation by reflection).

**A Computation and visualization of useful heuristic metrics on the individual, 3D-reconstructed head surface.**

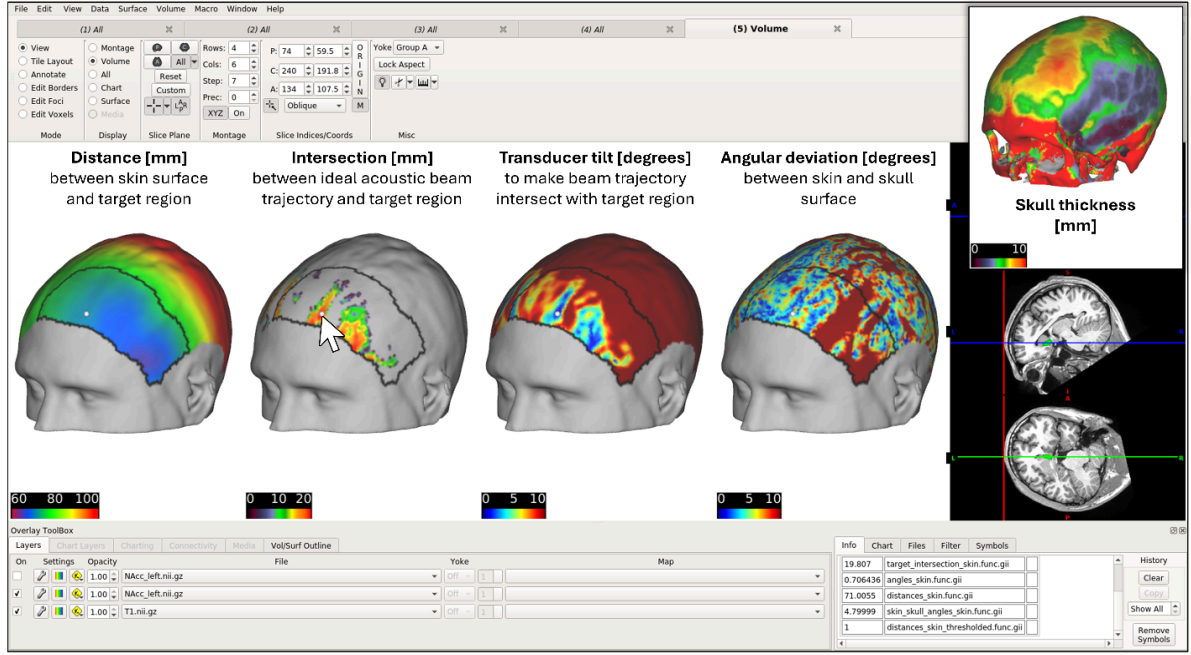

**B Illustration of a potential transducer placement and expected acoustic focus resulting from the position selected in A.**

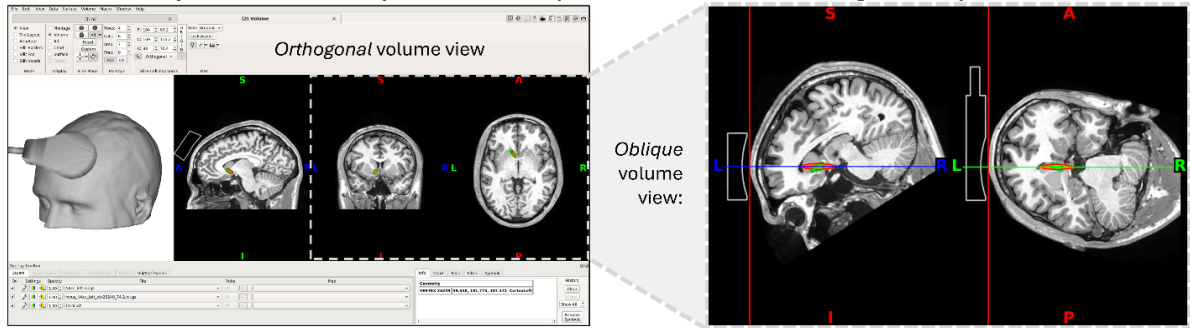

**C Export of selected transducer placement(s) for acoustic simulations and neuronavigation (using external software).**

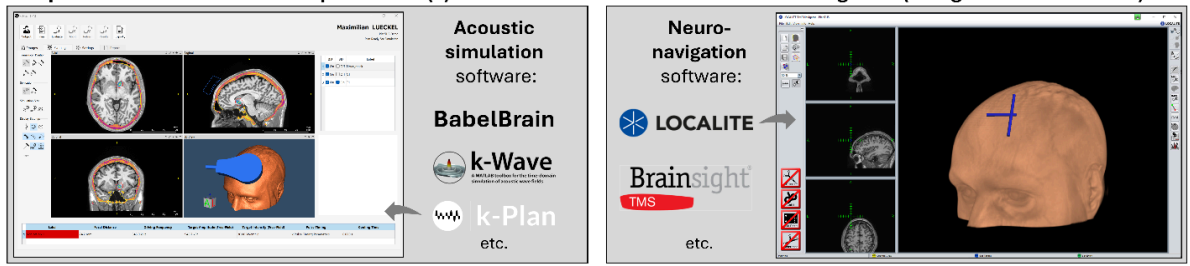

**D Review of acoustic simulation results.**

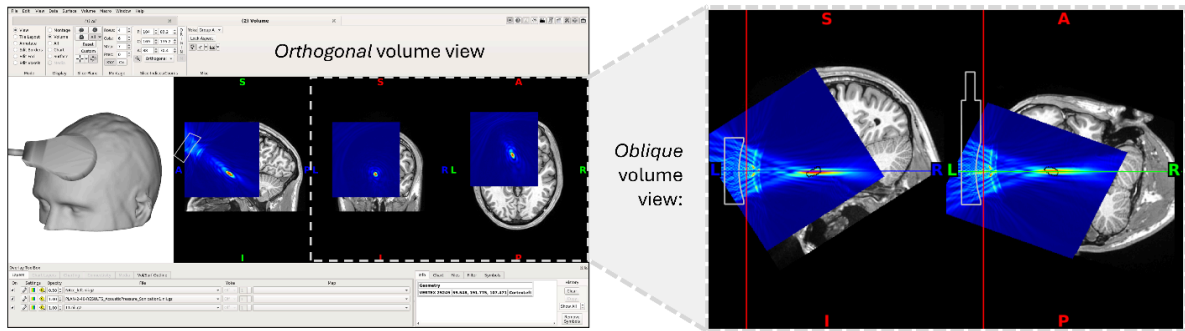

**Figure 1** Example workflow for the planning of TUS transducer placements using PlanTUS. **A**: Main graphical user interface, based on Connectome Workbench viewer. PlanTUS computes and visualizes

several useful heuristic metrics (for more details, see descriptions in the figure or in the main text) on the individual, 3D-reconstructed head surface that help to intuitively evaluate the most appropriate area(s) for transducer placement(s). By clicking anywhere on the head surface (illustrated by the computer mouse pointer icon), users can easily select the spot where they would like to place the (center of the) transducer. The volume view on the right allows to reorient the individual T1 image along a theoretical, idealized beam trajectory going straight from the selected spot on the head surface into the brain (blue and green lines in the upper and lower slice view panel, respectively; "Oblique" option in the "Slice interface/Coords" panel). This allows to provisionally evaluate the intersection between beam trajectory and target region (overlaid on the T1 image and visualized in green). The inlay in the upper right corner illustrates the visualization of individual skull thickness, as an additional output of PlanTUS. **B**: Once the user has selected and confirmed the desired transducer placement, another Connectome Workbench viewer window opens that (i) shows the potential transducer placement (including a simplified transducer model) on the 3D-reconstructed individual head surface (left part in the Connectome Workbench view on the left), and (ii) a volume view of the individual T1 image, with the mask of the target region (green) and a simplified, ideal acoustic focus (orange with red outline) overlaid (right part in the Connectome Workbench view on the left). Again, the T1 image can be reoriented along the theoretical, idealized beam trajectory going perpendicular from the transducer surface into the brain, aligning with the lateral axis of the simplified focus (zoomed-in oblique volume view on the right). **C**: Examples for the use of PlanTUS outputs. Left: Import of the transducer placement planned using PlanTUS in the k-Plan acoustic simulation software. Note that PlanTUS outputs are generally also compatible with other (open-source) acoustic simulation software, such as BabelBrain or k-Wave. Right: Import of the transducer placement planned using PlanTUS in the Localite neuronavigation software. Note that PlanTUS outputs are generally also compatible with other neuronavigation software, such as BrainSight. **D**: Example for the review of acoustic simulation results. Acoustic simulation results (in nifti format and in the same space as the T1 input image) can be easily visualized in the same environment (i.e., Connectome Workbench viewer), which helps to evaluate on- versus off-target coverage of the simulated acoustic focus (especially using the oblique volume view shown in a zoomed-in version on the right).

*3.2 Additional examples*

Figure 2A illustrates additional cases of individual transducer placement planning with PlanTUS.

For sonication of the thalamus (first row), extended areas on the head surface show large intersections of the simplified beam trajectory with the target, due to its comparatively large volume. Simulations resulting from the PlanTUS-selected transducer placement show a good overlap between acoustic focus and thalamus.

For sonication of the amygdala (second row), only a very small area over the temporal window shows strong intersection of the simplified acoustic beam trajectory and the target. Furthermore, the curvature of this area strongly deviates from that of the underlying skull (red areas in the right-most metric map). This indicates that the acoustic focus will likely be distorted by the skull – as confirmed by the corresponding acoustic simulation (note, however, that the simulated acoustic focus still overlaps with the target).

For sonication of the subgenual anterior cingulate cortex (third row), one might intuitively place the transducer laterally over the temporal bone window. However, PlanTUS suggests a more medial transducer placement over the forehead, which results in a good overlap of the (simulated) acoustic focus with the target.

Importantly, PlanTUS can also give insights into whether a brain region is targetable at all, given the specified properties of the transducer. The last row of Figure 2A illustrates that sonication of the ventromedial prefrontal cortex is not possible with a transducer that has no lateral steering capabilities (indicated by the lack of direct intersections between the simplified acoustic beam trajectory and target, and large angles of required transducer tilt).

Lastly, as illustrated in Figure 2B, PlanTUS can also be useful in early planning phases of TUS studies, before having acquired any individual MR images, for (i) testing the general feasibility of target exposure, or (ii) figuring out the technical requirements of a TUS system before it is being purchased or engineered/developed (e.g., minimum focal depth to reach a specific target region of interest, focal geometry to maximize on- and minimize off-target coverage), based on openly available anatomical MR images (from a representative sample such as the Human Connectome Project; Elam et al., 2021; Figure 2B, bottom) or even template images (e.g., in MNI standard space) (Figure 2B, top).

**A  Planning examples for different brain targets.**

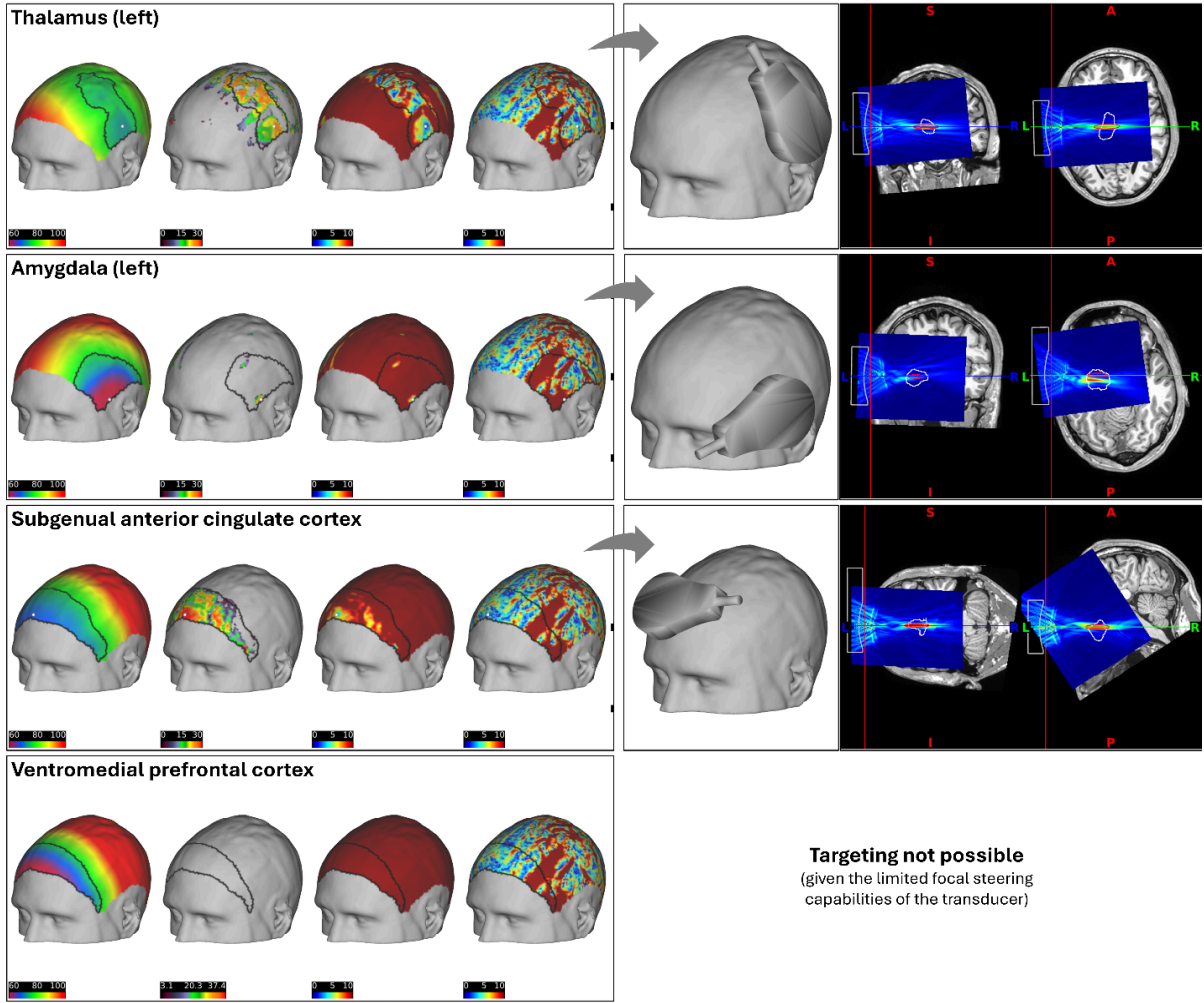

**B  Planning examples for different datasets.**

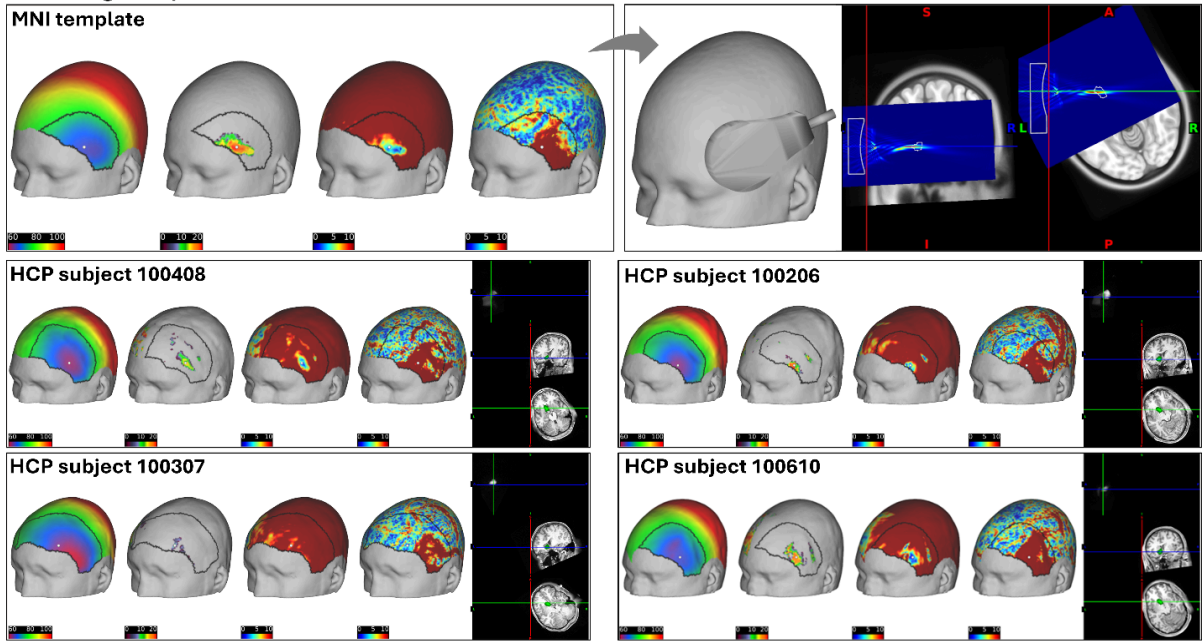

**Figure 2** Application examples of PlanTUS. **A**: Planning of transducer placements for sonicating different targets (left thalamus, left amygdala, subgenual anterior cingulate cortex, ventromedial

prefrontal cortex) in the same individual. The panels on the left illustrate the heuristic metrics that PlanTUS computes and visualizes on the individual, 3D-reconstructed head surface (from left to right: distance [mm] between skin surface and target region; intersection [mm] between ideal acoustic beam trajectory and target region; transducer tilt [degrees] to make beam trajectory intersect with the target region; angular deviation [degrees] between skin and skull surface). The right panel shows (i) the potential transducer placement resulting from the position marked in the left panel by the small white sphere as well as the simulated acoustic pressure field resulting from the respective transducer placement, reoriented along the transducer axes and overlaid on the individual T1 image (with a white outline of the target region and a red outline indicating the simplified, "expected" focus). **B**: Applications of PlanTUS in other, participant-independent datasets, for planning the sonication of the left Nucleus Accumbens (NAcc), e.g., in early planning phases of a TUS study. Upper row: Planning of transducer placement with PlanTUS using a template head/brain in MNI standard space. Note that the MNI head/brain is larger (higher, deeper, longer) than a normal head/brain (e.g., Lancaster et al., 2007), hence the area from which the target would be reachable given the maximum focal depth of the transducer (indicated by black outline on the head surfaces) will often be smaller in comparison to planning scenarios using individual MRI data (cf. Figure 1A or the lower four panels of Figure 2B), which should be taken into account when planning transducer placements based on the MNI template. Lower four panels: Planning of transducer placement with PlanTUS using four individual example datasets from the Human Connectome Project (HCP), illustrating considerable interindividual variability in where the transducer can and should be placed.

**4 Conclusion**

PlanTUS is an open-source tool that helps TUS users to address the recurring question of where to place the TUS transducer on the scalp of a given individual for optimally targeting a specific brain region of interest. It can also be used to generally assess the feasibility of targeting certain brain regions even before starting a TUS study. PlanTUS allows to interactively and heuristically select promising transducer placement(s), and its outputs are compatible with different acoustic simulation and neuronavigation software. Importantly, PlanTUS-selected transducer placements principally result in a very good overlap between the simulated acoustic focus and brain target. As such, PlanTUS is a versatile tool to support both new and experienced users of TUS in selecting optimal transducer placement(s).

**Acknowledgements**

T.O.B. received funding supporting this work from the Boehringer Ingelheim Foundation (grant on "Methods Excellence in Neurostimulation"), the European Innovation Council (EIC Pathfinder project CITRUS, Grant Agreement No. 101071008), the Ministry of Science and Health of the State of Rhineland-Palatinate, Germany (MWG "ACCESS" grant), the Leibniz Association (ScienceCampus "NanoBrain"), and the German Research foundation (DFG Grant No. 468645090).